\documentstyle[12pt]{article}

\newcommand{\beq}{\begin{eqnarray}}
\newcommand{\eeq}{\end{eqnarray}}
\makeatletter
\def\vereq#1#2{\lower3pt\vbox{\baselineskip1.5pt \lineskip1.5pt
\ialign{$\m@th#1\hfill##\hfil$\crcr#2\crcr\sim\crcr}}}
\makeatother

\begin{document}

\begin{titlepage}
\begin{center}
    \hfill    LBNL-43586 \\
{}~{} \hfill UCB-PTH-99/28  \\
{}~{} \hfill hep-ph/9906513\\
\vskip .3in
{\Large \bf Cosmology of One Extra Dimension \\ with Localized Gravity}

\vskip 0.3in
{\bf Csaba Cs\'aki\footnote{Research fellow, Miller Institute for
Basic Research in Science.}, Michael Graesser, Christopher Kolda, and
John Terning}

\vskip 0.15in

{\em Theoretical Physics Group\\
     Ernest Orlando Lawrence Berkeley National Laboratory\\
     University of California, Berkeley, California 94720}

\vskip 0.1in
\vskip 0.1in

{\em Department of Physics\\

    University of California, Berkeley, California 94720}

\vskip 0.1in
{\tt  ccsaki@lbl.gov, mlgraesser@lbl.gov, \\
cfkolda@lbl.gov,
terning@alvin.lbl.gov }

\end{center}

\vskip .25in
\begin{abstract}
We examine the cosmology of the two recently proposed scenarios for a
five dimensional universe with localized gravity. We find that the scenario
with a non-compact fifth dimension is potentially viable, while the scenario
which might solve the hierarchy problem predicts a contracting
universe, leading to a variety of cosmological problems.
\end{abstract}
\end{titlepage}

\newpage
\setcounter{footnote}{0}

The main theme of the first 20 years of the hierarchy problem has been
to modify particle physics around the TeV scale. In the past two years
it has become evident that another successful route to solving the
hierarchy problem is to modify the nature of gravitational interactions
at distances shorter than a millimeter~\cite{HW,Nima,otherextra}.
This modification can be most simply achieved
by introducing compact extra dimensions. One recent proposal which attracted
enormous attention is to lower the fundamental Planck scale $M_*$ all the
way to the TeV scale, by introducing large extra dimensions~\cite{Nima}. 
The observed
Planck scale is then just an effective scale valid for energies below the
mass scale of the 
Kaluza-Klein (KK) excitations. The consequence of this proposal is
that the necessary size of the extra dimensions is determined by
the KK reduction formula
\begin{equation}
\label{eq:KK}
R=\left( \frac{M_{Pl}}{M_*}\right)^{\frac{2}{n}} \frac{1}{M_*},
\end{equation}
where $M_*$ is the fundamental Planck scale of the order of
1 TeV, $M_{Pl}=10^{18}$ GeV, and $n$ is the number of extra dimensions.
Applying this formula for one extra dimension one obtains
$R\sim 10^{13}$ cm, which would immediately suggest that
this possibility is excluded because gravity would be modified
at the scale of our solar system. For $n\geq 2$, $R$ is sufficiently small
so that this scenario is not excluded by
short-distance gravitational measurements. However, Randall and Sundrum
recently realized \cite{randall,randall2} 
that the case of one extra dimension is very special,
and the na\"{\i}ve KK reduction formula (\ref{eq:KK}) may not be applicable in
this case. The basic reason behind this is that in this scenario the
standard model fields have to be confined to a three-dimensional
wall (``three-brane''), and such branes act like sources for gravity in the 
extra
dimensions. The behavior of  Green's functions in one dimension is
dramatically different from the case of two or more dimensions.
Indeed, in one dimension Green's functions grow linearly, while
in the case of more than two dimensions there is an inverse power law
$1/r^{d-2}$ (and  logarithmic growth for $d=2$). Thus
branes act like  small perturbations on the system for
the case of two or more extra dimensions, and one expects the KK reduction
formula (\ref{eq:KK}) to be applicable. For just one extra dimension, however,
the presence of branes can significantly alter the bulk gravitational fields,
which may invalidate arguments based on the na\"{\i}ve KK reduction formula.

Indeed, Randall and Sundrum (RS) have presented a new static classical solution
to Einstein's equations with one extra dimension (taken to be $S^1/Z_2$),
and branes with non-vanishing tensions placed at the orbifold fixed
points~\cite{randall}. In this solution, for large brane separations,
the effective four-dimensional Planck scale
is independent of the size of the extra dimension, in agreement with the
expectation that (\ref{eq:KK}) is invalid for the case of one extra
dimension. For their solution to work RS found that the brane tensions must
be of opposite sign, and that there must be a negative bulk cosmological
constant
which
stabilizes the system.

Two possible applications of this solution have been
proposed~\cite{randall,randall2}. In one case (which we refer to as RS1),
our universe (``the visible brane'') is the brane with negative tension,
and the exponential ``warp-factor'' appearing in the RS solution will yield
a natural new solution to the hierarchy problem~\cite{randall}.
In the second (RS2) proposal~\cite{randall2},
the visible brane is the one with positive tension. In this case the
hierarchy problem is not solved, however, the second brane can be moved
to infinity, thus providing an exciting example of a non-compact extra
direction~\cite{noncompact}, 
which nevertheless correctly reproduces Newton's law
on the visible brane.

Bin\'etruy, Deffayet and Langlois (BDL), however, have pointed out
recently~\cite{binetruy}
that five dimensional theories with branes tend to have non-conventional
cosmological solutions, once matter on the walls is included.\footnote{For
other recent results on the cosmological aspects of theories
with large extra dimensions see [8-16].}
In this letter, we apply the results of BDL to the models presented
by RS. We find that the equation for the
scale factor on the visible brane (for small matter densities) coincides with
the
conventional Friedmann equation, up to the overall
sign of the source terms. This sign
depends on the sign of the cosmological constant (tension) on the visible
brane.
In the
case of negative brane tension the source terms of the Friedmann-like equation
have the opposite sign from standard cosmology.
The change in sign implies that the universe
would collapse on a timescale on the order of the Hubble time
at the start of the expansion,
for any matter with energy density $\rho$, pressure
$p$, on our wall with an equation of state $p=w \rho$
and $w<1/3$.
So whereas in the radiation-dominated phase the universe expands as in
the conventional cosmology, after the transition from the radiation-dominated
to matter-dominated (or quintessence-dominated) epoch when the
universe was a few thousand years old, the universe would
not expand as in conventional cosmology, but would rather
collapse within a few thousand years.
As we argue, this conclusion relies on knowing the expansion
rate during the era of matter-radiation equality, which is
provided by the success of the standard big-bang nucleosynthesis
(BBN), and the current baryon and radiation densities. So in order to
avoid this conclusion this scenario requires either a non-standard
BBN or a modification of the RS1 solution, for example through the 
introduction of additional fields.
We emphasize that crucial to these
conclusions is the fact that we live on a negative-tension
brane, and that there is only one extra dimension.  Both of these
facts, however, are also crucial ingredients to the
solution to the hierarchy problem presented by RS.
For the case of the positive brane tension, however,
the conventional expanding solution is reproduced in this model.

We begin by summarizing the work of BDL \cite{binetruy}.
The scenario considered here is a five-dimensional spacetime
compactified on the line segment $S^1 / Z_2$. The bulk
coordinate is labeled by $y$, which is taken to be in the interval
$-1/2 \leq y \leq 1/2$, where the points $-1/2$ and $1/2$ are identified.
In this notation  the coordinate $y$ is dimensionless.
The $Z_2$ symmetry identifies the
points $y$ and $-y$, so we can restrict ourselves to $0 \leq y \leq 1/2$.
Two three-branes are placed at the fixed points of the discrete symmetry,
$y=0$ and $y=1/2$. In our notation, the visible brane
is located at
$y=0$, and a hidden brane is located at $y=1/2$.
The compactness of the extra dimension requires the existence of
two branes, since each brane must absorb
the gravitational flux lines from the other.

The most general metric for a five-dimensional spacetime
which preserves three-di\-men\-sio\-nal rotational and translational
invariance is given by
\begin{equation}
ds^2= n^2(\tau,y) d \tau^2 -a^2(\tau,y)d \vec{x}^2 -b^2(\tau,y) dy^2.
\label{metric}
\end{equation}
Note that here we use the metric signature $(+,-,-,-,-)$. The
induced metric on the visible brane is obtained by evaluating the metric
tensor at $y=0$. The
metric is determined by solving Einstein's equations in the presence
of some energy density in the bulk and on the three-branes. A solution
which describes a static four-dimensional Lorentz invariant
universe is given by $a(\tau,y)=n(\tau,y)=f(y)$. This still
allows for a non-trivial
dependence of the metric on the bulk coordinate. This is the key point
in the solution of \cite{randall} to the hierarchy problem,
for localizing gravity in the bulk \cite{randall,randall2}, and for
obtaining a non-compact fifth dimension with a conventional
Newton's force law \cite{randall2}. For
an expanding four-dimensional universe, however, we
must have $a \neq n$.
The metric is determined by solving Einstein's equations
\begin{equation}
G_{AB}\equiv R_{AB} -\frac{1}{2} g_{AB} R = \kappa^2 T_{AB}.
\end{equation}
Here $A$, $B= 0,1,2,3,5$, and $\kappa^2$ is the five-dimensional
Newton's constant, which is related to the five-dimensional Planck scale
by $M^3_* = \kappa ^{-2}$. The components of the Einstein tensor
$G_{AB}$ with the above ansatz are given in (8)--(11) of \cite{binetruy}.

The five-dimensional stress-energy tensor
$T_{AB}$ is the sum of contributions from the
bulk and from the two branes.
The width of the branes is neglected since it is
$O(1/M_*)$, which is much smaller than the distance scales of
cosmological interest. Thus the
stress-energy tensor is approximated as
\begin{equation}
T^{AB} (x,y)=\tilde{T}^{AB}(x,y) + \frac{S_{vis}^{AB}(x)}{b_0(\tau)}
\delta(y)
 +\frac{S_{hid}^{AB}(x)}{b_{1/2}(\tau)}\delta(y-1/2),
\end{equation}
where $b_0(\tau) \equiv b(\tau,0)$ and $b_{1/2}(\tau)\equiv b(\tau,1/2)$.
Here $\tilde{T}^{AB}$ is the stress-energy tensor in the bulk,
and $S_{vis}^{AB}$ ($S_{hid}^{AB}$) is the 4-dimensional stress-energy tensor
of the visible (hidden) brane: 
${{S}^A _B}_{vis} =(\rho, -p, -p, -p,0)$ and
${{S}^A _B}_{hid} =(\rho _*, -p_*,-p_*,-p_*,0)$. At this point the
composition of the energy density on the walls is completely general.

Using the above energy-momentum tensor BDL derived an equation for the
scale factor on the visible brane which is independent of the details
of the global solution to Einstein's equations\footnote{This equation
can be derived \protect\cite{binetruy}
by first calculating the discontinuities in the derivatives 
of the
functions $a$ and $n$ at the position of the branes by matching the
$\delta$--functions in Einstein's equations. This information about the
discontinuity of the derivatives together with the discontinuity and the 
average value of the
5,5 component of Einstein's equation results in (\protect\ref{FRWlike}).}:
\begin{equation}
\frac{\ddot{a}_0}{a_0} + \left( \frac{\dot{a}_0}{a_0}\right)^2 =
-\frac{\kappa^2}{3 b^2 _0} \tilde{T}_{55} -
\frac{\kappa^4}{36} \rho (\rho + 3 p) .
\label{FRWlike}
\end{equation}
The time derivatives which appear are with respect to $t$, where
$dt \equiv n(\tau,0) d \tau$, and $a_0(t)= a(\tau,0)$ is the
scale factor on the visible brane.
There are several important features of this equation. First, the
energy density and pressure of the second brane do not appear. This follows
from the local nature of Einstein's equations~\cite{Israel}. 
Second, as noted in
\cite{binetruy}, the Hubble parameter $H \equiv \dot{a}_0/ a_0 \propto
\rho$ rather than $\sqrt{\rho}$ as in the conventional cosmology. Finally,
this
equation depends only on $a_0$. In the case that $\tilde{T}^5 _5$
is time-independent (as will be the case later), this equation
completely determines $a_0(t)$. Therefore it is not
necessary to determine the solutions to
Einstein's equations for the whole bulk.

We assume that no energy is flowing from the
bulk onto the brane so that $T^{05}=0$. Then
the discontinuity in the $G_{05}=0$ equation \cite{binetruy}
gives the usual conservation of
energy condition:
\begin{equation}
 \dot{\rho} +3 \frac{\dot{a}_0}{a_0} (\rho +p) =0.
\label{conserve}
\end{equation}
Thus for $p=w \rho$, $\rho \propto a^{-3( w+1)}$ as in standard
cosmology.  The conservation of energy equation (\ref{conserve}) and
the Friedmann-like equation (\ref{FRWlike}) are the central equations
which are used in what follows.

Now we briefly review the solution  presented by Randall and
Sundrum \cite{randall, randall2}.
In their scenario the two branes have some tension, $V_{vis}$,
$V_{hid}$, and the bulk contains a cosmological constant $\Lambda$.
In terms of our earlier notation,
$\rho=-p=V_{vis}$, $\rho_*=-p_*=V_{hid}$, and
$\tilde{T}^A _B = \Lambda (1,1,1,1,1)$. A static Lorentz invariant
solution is obtained with $a(\tau,y)=n(\tau,y)= e^{\sigma(y)}$, and
$b(\tau,y)=b(y)=b_0=$ const., the latter having been obtained by a coordinate
transformation on $y$. Then the 5,5 component of Einstein's equation gives
\begin{equation}
 {\sigma'} ^2 = -\frac{\kappa^2 b^2 _0}{6} \Lambda.
\end{equation}
Thus $\Lambda <0 $ is required. It is then convenient to introduce
\begin{equation}
 m ^2 \equiv -\frac{\kappa^2}{6} \Lambda.
\label{rel1}
\end{equation}
Therefore $\sigma = \pm m b_0 |y|$. (The absolute value is required so
that $a''$ is singular at $y=0$, matching the $\delta$--function sources in
Einstein's equations.)
The discontinuity equation for either $a$ or $n$
then requires that
\begin{equation}
 V_{vis}=\mp \frac{6}{\kappa^2} m =-V_{hid}.
\label{rel2}
\end{equation}
We note that
the case in which $V_{vis} \neq -V_{hid}$ leads to the brane inflating
solutions found in \cite{kaloper}. Since we are 
interested in non-inflationary solutions, we assume that
the tensions of the two branes are adjusted to satisfy
(\ref{rel2}). Since $sgn(V_{vis})$ is crucial in determining
the expansion rate of our wall discussed below, we emphasize
that the correlation between the sign
of the tension of the visible brane  and the growth of the scale
factor away from the visible brane is:
\begin{equation}
{\rm RS1:} \hbox{ } n(y)=a(y)=e^{+ m b_0 |y|}  \Longleftrightarrow V_{vis}
<0,
\end{equation}
\begin{equation}
{\rm RS2:} \hbox{ }n(y)=a(y)=e^{- m b_0 |y|}  \Longleftrightarrow V_{vis}
>0.
\end{equation}
For RS2, gravity is localized about our brane \cite{randall2}.
This allows for a non-compact fifth dimension that is consistent
with the short-distance force experiments. By contrast, case RS1 with
$V_{vis}<0$ provides a potential solution to the hierarchy problem
\cite{randall}.
The
reason for this is the following. The metric on the distant brane
contains a conformal factor $e^{m b_0}$ in units where the conformal
factor on our wall is one. This implies
that mass scales on our wall
and the distant wall are then related by $m_{vis} = e^{-m b_0/2} m_{hid}$,
so that a large hierarchy of mass scales is possible if
$m_{hid} \sim M_{Pl} \sim M_*$ and $m b_0 \sim 100$.

This non-trivial scale factor significantly modifies the na\"{\i}ve
relation $M^2 _{Pl} \sim  M_{*} ^3 b_0 $ between the fundamental and
derived Planck scales. The correct relation for
$m b_0 \gg 1$ is in fact~\cite{randall}
\begin{equation}
8 \pi G_N = \frac{1}{M^2_{Pl}} = \kappa^2 m
= \frac{\kappa^4}{6} |V_{vis}|.
\label{plmass}
\end{equation}
It is remarkable that this is independent of the size of the extra
dimension.  For RS1
the coupling of the KK excitations of the graviton to matter are given
by $\sim e^{m b_0 /2} /m$ \cite{randall2}. Since these excitations
would appear as resonances in collider experiments, this coupling
must be $O($TeV$^{-1})$ or smaller, thus implying $m \sim M_{Pl}$
(since $e^{mb_0/2}\sim 10^{15}$). Therefore a satisfactory resolution  to the
hierarchy problem requires
$m \sim \kappa^{-2/3} \sim M_{Pl}$.

However, both of these solutions to Einstein's equations are static
and do not describe a universe with time-dependent scale
factor $a$. While some work has been done on inflating 
solutions~\cite{kaloper}, we
will attempt to uncover cosmological solutions which reproduce the
successes of the usual Friedmann equation for a flat universe. 
This seems difficult at
first thought, given the earlier statement that in
five-dimensional brane models $H\propto \rho$. However, we will see that the 
presence of large background cosmological constants changes this
conclusion, and the usual Friedmann equation is reproduced, up to
the sign of the source term.

We begin by perturbing the RS solution by placing an additional energy 
density on 
the two branes without a compensating change to the bulk cosmological constant.
That is, we consider $\rho=V_{vis}+\rho_{vis}$ and $p=-V_{vis}+p_{vis}$ where 
$V_{vis}$ is given by (\ref{rel2}) and $\rho_{vis}$, $p_{vis}$ are the 
energy density and pressure measured by an observer living on the visible 
brane, with equation of state $p_{vis}=w \rho_{vis}$. 
Key to our results will be that 
we work in the limit 
$\rho_{vis}\ll |V_{vis}|$. Given that $|V_{vis}|\sim M_{Pl}^4$ in these models,
this limit is the correct one for describing our (post-inflationary)
universe.
Substituting these expressions for $\rho$ and $p$ into (\ref{FRWlike})
gives, for either RS1 or RS2,
\begin{equation}
\frac{\ddot{a}_0}{a_0} + \left( \frac{\dot{a}_0}{a_0}\right)^2 =
-\frac{\kappa^4}{36} V_{vis}(3p_{vis}-\rho_{vis}) -\frac{\kappa^4}{36} \rho_{vis}
(\rho_{vis} +3 p_{vis})   .
\label{FRW2}
\end{equation}
The $O(\Lambda)$ and $O(\kappa^4 V_{vis} ^2)$ terms cancel using
(\ref{rel1}) and (\ref{rel2}).
Note that the presence of the background energy density allows for
$H^2 \propto \rho_{vis}$ as in conventional cosmology. This
differs from the observations of BDL because here the brane matter is a
perturbation to the background RS solution. It is also
clear that
the presence of the prefactor $V_{vis}$ in (\ref{FRW2})
implies that  for one
of RS1 or RS2 solutions, (\ref{FRW2}) will have a negative sign relative to the
conventional Friedmann equation. In fact, we find that the ``wrong-signed''
Friedmann-like equation corresponds to RS1,
the solution with $V_{vis} <0$. To see this,
substitute the formula for the Planck mass,
(\ref{plmass}), into (\ref{FRW2}):
\begin{equation}
{\rm RS1}:  \hbox{    }
\frac{\ddot{a}_0}{a_0} + \left( \frac{\dot{a}_0}{a_0}\right)^2 =
\frac{4 \pi G_N}{3}( 3p_{vis} -\rho_{vis}) -\frac{\kappa^4}{36} \rho_{vis}
(\rho_{vis} +3 p_{vis})   .
\label{feqn}
\end{equation}
For small densities
($\rho_{vis} \ll M^4 _{Pl}$) we can neglect the second term on
the RHS, since $\kappa^4\sim 1/M_{Pl}^6$.
The first term on the RHS contains a negative sign relative to the
Friedmann equation in the conventional cosmology.
Obviously this sign is flipped in the RS2 scenario and
we get the correct Friedmann equations up to small corrections.

It is interesting to note that for a radiation-dominated (RD) universe this
sign problem has no effect: since $p_{vis}= \rho_{vis}/3$ the first term
on the RHS of (\ref{feqn}) identically vanishes, and one obtains the 
same equation for both RS1 and RS2. It is then important to see
whether the conventional RD solution is obtained in this case.
  The approximate
solution for the scale factor can be easily found. We take
\beq
\rho_{vis}(t) = \rho_{vis} (t_i)
\left({{a_0(t)}\over{a_0(t_i)}}\right)^{-4} ~,
\eeq
then
\beq
\label{RD}
a_0(t) \sim t^{{1}\over{2}}\left(1-{{\kappa^4}\over{36}} {{
\rho_{vis}^2(t_i)
a_0(t_i)^8}\over{t^2}} +\ldots \right) ~.
\label{RDsol2}
\eeq
Thus at late times this reduces to the evolution of standard RD
cosmology. Note, that this solution is different from the
$a_0(t) \sim t^{1/4}$ solution presented in \cite{binetruy} (even though
(\ref{FRWlike})
reduces to exactly the same equation in the case of a RD universe).
The reason is simply that a non-linear second order differential
equation can have more than one solution, and the initial condition
will determine which is the relevant one. Since (\ref{RD}) reproduces the
standard RD cosmology, it is plausible that the potential problems with
BBN found in \cite{binetruy} are in fact
solved by the existence of the
extra solution (\ref{RD}). From the solution (\ref{RDsol2}) we can see that 
both RS1 and RS2 reproduce the conventional RD solution. Since 
for RS2 (\ref{FRW2}) gives the conventional Friedmann equation 
for any type of matter (up to
small corrections), we conclude that the RS2 solution is viable. 
Therefore in the following we will only concentrate on the RS1 solution.

What is the effect of the wrong sign in the Friedmann equation for the
RS1 solution? Assume 
that after some time $t_{eq}$ the energy density of
the universe is dominated by a component
with an equation of state having $w <1/3$.
In what follows, the subscript ``$eq$'' will denote quantities measured
at  time $t_{eq}$ of matter-radiation equality. Then, e.g.,
$\rho_{vis}(t) = \rho_{eq} (a_0(t)/a_0(t_{eq}))^{-3(w+1)}$.
Next, for $w \neq 1/3$ and energy densities $\rho_{vis} \ll M^4 _{Pl}$,
the second term on the RHS of (\ref{feqn}) is subdominant
to the first, so it is neglected.
It is then convenient to
introduce new variables $ \tilde{t} \equiv H_{eq} t$ and
 $x(\tilde{t}) \equiv a_0(t)/ a_0(t_{eq})$, where $H(t)$ is the Hubble 
parameter.
The initial conditions at
$t=t_{eq}$ are then $x(t_{eq})=1$ and $\dot{x}(t_{eq}) =1$, where
the overdot denotes a derivative with respect to $\tilde{t}$.
Then
in these units (\ref{feqn}) is
\begin{equation}
\frac{\ddot{x}}{x}+ \left( \frac{\dot{x}}{x} \right)^2=-
\frac{\lambda}{2} \frac{1}{x^{3(w+1)}},
\end{equation}
where the dimensionless constant
$\lambda \equiv 8 \pi G_N  \rho_{eq} (1-3 w)/(3 H^2_{eq})$
is introduced. Note that for $w<1/3$ one has $\lambda>0$.
(One obtains conventional cosmology
by the replacement $\lambda \rightarrow - \lambda$.)
Also note that if $H^2 = 8 \pi G_N \rho/3$ were the
correct flatness constraint equation then $\lambda=1-3w$. Since
it is expected that
$H^2 \sim G_N \rho$, then $\lambda \sim O(1)$.
Multiplying the above equation by $y \equiv x^2$ then gives
\begin{equation}
\ddot{y} =- \lambda y^{-(3w+1)/2}.
\label{leqn}
\end{equation}
That is, the expansion of our universe is described by the
one-dimensional motion of
a particle in the
classical potential
\beq
V(y) = \frac{2}{1-3w} \lambda y^{(1-3w)/2}.
\label{newton}
\eeq
For $w<1/3$ this describes
an {\em attractive} potential. From this we understand that the
universe reaches a maximum size, and then collapses. This
is one of the main points of this letter: this result is in conflict with our
current understanding of cosmology from several viewpoints.
At the very least,
observationally the universe seems to be accelerating \cite{Perlmutter}
rather than decelerating. (We will examine
a sharper conflict with BBN and the age of the universe below.)
By contrast, in
conventional cosmology there is an extra negative sign
on the RHS of (\ref{leqn}), so for $w<1/3$ the potential
is inverted and the universe expands forever (since curvature terms
have not been included). In the RS1 scenario, the
period of this oscillation is $O(1)$ in these dimensionless
units if $\lambda \sim O(1)$. In fact, for $w=0$ (non-relativistic matter),
the solution is
\begin{equation}
 \tilde{t}= \int \frac{x dx}{\sqrt{1+  \lambda - \lambda x}}
=\frac{2}{\lambda^2}
\left(\frac{1+  \lambda- \lambda x}{3}-1-\lambda \right)
\sqrt{1+\lambda - \lambda x}.
\end{equation}
In RS1, the period $t_U$ of oscillation (i.e., the age of the universe)
for $\lambda=1$
is then $t_U =20/3$. So in the original units
$t_U \sim H^{-1} _{eq} $ as expected. By inspection the
maximum size of the universe is $x_{max}=1 + 1/\lambda$ for $w=0$.
Finally, note that for the conventional
MD cosmology ($\lambda=-1$) the
above formula correctly reproduces $\tilde{t} \sim  x^{3/2}$.

In the  RS1 scenario, the classical potential (\ref{newton})
determines a first-order equation for $x$, in other words, it
determines $H^2= H^2 _{eq} \dot{x}^2 / x^2$. In fact,
\begin{equation}
\label{Hubble}
H^2 =\frac{EH^2_{eq}}{2 x^4}- \frac{8 \pi G_N}{3} \rho_{vis}.
\label{funnyflatness}
\end{equation}
This contains an arbitrary integration constant $E=2 +2 \lambda/(1-3w)$.
Note that the above equation (\ref{Hubble}) together with the 
conservation of energy (\ref{conserve}) implies the Friedmann equation
(\ref{feqn}) for arbitrary values of the integration constant $E$.
In conventional cosmology $\lambda =-1+3w$ so $E=0$, and there
is an obvious extra minus sign in the second term, reducing 
(\ref{funnyflatness})
to the usual flatness equation. In RS1 (\ref{funnyflatness}) is
the analog of the flatness equation. In what follows,
we assume that $\rho_{vis}>0$, which implies
\beq
E > {{16 \pi G_N \rho_{eq} }\over{ 3 H^2_{eq}}} >0 ~.
 \eeq

The preceding arguments indicate that in the RS1 scenario,
after the time of matter-radiation equality, the universe collapses
on a timescale given by $H^{-1}_{eq}$.
The expansion rate $H_{eq}$ is obtained as follows. Assuming standard
BBN, we know the radiation temperature $T_{BBN} \sim $ MeV, 
and expansion rate $H_{B\!B\!N}$ during BBN. We use the
present-day values of the radiation and baryon energy densities,
together with conservation of energy and (\ref{RDsol2})
describing the RD cosmology, to
determine $H_{eq}$. We use $a \sim t^{1/2}$ 
from (\ref{RDsol2})
rather than the  solution $a \sim t^{1/4}$, 
which is known to have difficulties with the
He$^4$ abundance \cite{binetruy}.
Then
\begin{equation}
\label{Heq}
\frac{T_{eq}}{T_{B\!B\!N}}=\frac{a_{B\!B\!N}}{a_{eq}}=
\left(\frac{t_{B\!B\!N}}{t_{eq}}\right)^{1/2}
=\left(\frac{H_{eq}}{H_{B\!B\!N}}\right)^{1/2}.
\end{equation}
The first equality follows from
energy conservation, and 
the last two equalities from (\ref{RDsol2}). 
Next, conservation of energy is used, together with
the present-day value
$\rho_{\gamma} / \rho_{crit} \sim 10^{-5}$, and $T_{now} \sim 2.7 $ K,
to give $T_{eq} \sim 5$ eV. Inserting this result, together
with $T_{B\!B\!N} \sim $ MeV, and $H_{B\!B\!N} \sim T^2_{B\!B\!N}/M_{Pl} \sim
10^{-21}$ MeV, into (\ref{Heq}) gives $H_{eq} \sim 10^{-32}$ MeV
(the standard BBN result). Therefore,
a standard BBN cosmology implies in the RS1 scenario that
an MD universe collapses on a time scale of
$t_{U} \sim H^{-1}_{eq} \sim {\rm few} \times 10^3$ years. It is clear
from these arguments that the age of the universe has not been
used as an input, so this last result may be viewed as the
RS1 standard BBN cosmology prediction for the age of the universe.

In order to evade the previous arguments RS1 requires some
non-standard version of nucleosynthesis. In particular, in order for
the universe to exist for billions of years, the Hubble parameter
at the start of MD must be orders of magnitude larger than in standard BBN:
$H^2_{eq} \sim 10^3 \times (\frac{8\pi}{3} G_N \rho_{vis})$.

~

To conclude, 
we have considered the cosmology of five-dimensional theories with
localized gravity on a three-brane introduced in~\cite{randall,randall2}. 
We have found that the solution with a non-compact extra dimension is
potentially viable, since it reproduces the conventional cosmological 
solutions. However, the solution which may solve the hierarchy problem
predicts a contracting universe with a lifetime of a few thousand
years (assuming standard BBN). One way to 
avoid this difficulty might be a non-conventional
BBN. But it seems more likely that these problems could be 
avoided by introducing additional matter fields in the bulk (which are
anyway required to stabilize the radius of the extra dimension). The 
difficulty with this possibility will be maintaining the features 
which led to the solution of the hierarchy problem in the first place.


\section*{Acknowledgements}
We are grateful to Nima Arkani-Hamed, Gia Dvali, and Nemanja Kaloper
for useful discussions and to Lisa Randall for comments on the manuscript. 
This work was
supported in part by the U.S. Department of Energy under Contract
DE-AC03-76SF00098 and in part by the National Science Foundation under
grant PHY-95-14797. C.C. is a research fellow of the Miller Institute
for Basic Research in Science.


\begin{thebibliography}{99}

\bibitem{HW}
P.~Horava and E.~Witten,
Nucl. Phys. {\bf B475}, 94 (1996);
Nucl. Phys. {\bf B460}, 506 (1996).

\bibitem{Nima}
N.~Arkani-Hamed, S.~Dimopoulos and G.~Dvali,
Phys. Lett. {\bf B429}, 263 (1998);
I.~Antoniadis, N.~Arkani-Hamed, S.~Dimopoulos and G.~Dvali,
Phys. Lett. {\bf B436}, 257 (1998).

\bibitem{otherextra}
I.~Antoniadis,
Phys. Lett. {\bf B246}, 377 (1990);
J.D.~Lykken,
Phys. Rev. {\bf D54}, 3693 (1996);
R.~Sundrum,
Phys. Rev. {\bf D59}, 085009 (1999);
K.R.~Dienes, E.~Dudas and T.~Gherghetta,
Phys. Lett. {\bf B436}, 55 (1998);
G.~Shiu and S.H.~Tye,
Phys. Rev. {\bf D58}, 106007 (1998);
Z.~Kakushadze and S.H.~Tye,
Nucl. Phys. {\bf B548}, 180 (1999).



\bibitem{randall}
L.~Randall and R.~Sundrum,
hep-ph/9905221.


\bibitem{randall2}
L.~Randall and R.~Sundrum, hep-th/9906064.

\bibitem{noncompact}
V.A.~Rubakov and M.E.~Shaposhnikov,
Phys. Lett. {\bf 125B}, 139 (1983);
M.~Visser,
Phys. Lett. {\bf 159B}, 22 (1985);
E.J.~Squires, Phys. Lett. {\bf 167B}, 286 (1986);
M.~Gogberashvili, hep-ph/9812296;
hep-ph/9812365;
hep-ph/9904383.

\bibitem{binetruy}
P.~Binetruy, C.~Deffayet and D.~Langlois,
hep-th/9905012.

\bibitem{kaloper}
N. Kaloper, hep-th/9905210; T.~Nihei, hep-ph/9905487.

\bibitem{Ovrut}
A.~Lukas, B.A.~Ovrut and D.~Waldram, hep-th/9806022;
hep-th/9902071.

\bibitem{Nimacosm}
N.~Arkani-Hamed, S.~Dimopoulos and G.~Dvali,
Phys. Rev. {\bf D59}, 086004 (1999);
N.~Arkani-Hamed, S.~Dimopoulos and J.~March-Russell,
hep-th/9809124;
N.~Arkani-Hamed, S.~Dimopoulos, N.~Kaloper and J.~March-Russell,
hep-ph/9903224;
C.~Cs\'aki, M.~Graesser and J.~Terning, hep-ph/9903319.

\bibitem{Dvali}G.~Dvali and S.H.~Tye,
Phys. Lett. {\bf B450}, 72 (1999);
G.~Dvali and G.~Gabadadze, hep-ph/9904221;
G.~Dvali and M.~Shifman, hep-th/9904021;
G.~Dvali, hep-ph/9905204.

\bibitem{banks}
T.~Banks, M.~Dine and A.~Nelson,
JHEP {\bf 06}, 014 (1999);
T.~Banks,
hep-th/9906126.

\bibitem{halyo}
E.~Halyo, hep-ph/9905244;
hep-ph/9905577.

\bibitem{other1}
D.H.~Lyth,
Phys. Lett. {\bf B448}, 191 (1999);
N.~Kaloper and A.~Linde,
Phys. Rev. {\bf D59}, 101303 (1999);
J.M.~Cline, hep-ph/9904495.

\bibitem{other2}
K.R.~Dienes, E.~Dudas, T.~Gherghetta and A.~Riotto,
Nucl. Phys. {\bf B543}, 387 (1999);
A.~Mazumdar, hep-ph/9902381;
P.~Kanti and K.A.~Olive, hep-ph/9903524;
A.~Pilaftsis, hep-ph/9906265.

\bibitem{other3}
S.~Cullen and M.~Perelstein, hep-ph/9903422;
H.A.~Chamblin and H.S.~Reall, hep-th/9903225;
A.~Riotto, hep-ph/9904485.
L.J.~Hall and D.~Smith,
hep-ph/9904267.


\bibitem{Israel}
W.~Israel, Nuovo Cim. {\bf B44}, 1 (1966). 

\bibitem{Perlmutter}
A. Riess et. al., Astroph. J. {\bf 116}, 1009 (1998); S. Perlmutter et. al.,
astro-ph/9812133.

\end{thebibliography}
\end{document}